\documentclass{sig-alternate}

\usepackage[
  pass,
]{geometry}

\usepackage[T1]{fontenc}
\usepackage{microtype}

\usepackage[pdftex]{hyperref}
\usepackage{url}      
\usepackage{cleveref}
\usepackage{float}
\usepackage{balance}

\usepackage{tabularx}
\usepackage{booktabs}
\newcommand{\ra}[1]{\renewcommand{\arraystretch}{#1}}
\newcommand{\specialcell}[2][c]{%
  \begin{tabular}[#1]{@{}r@{}}#2\end{tabular}}
  \newcolumntype{R}[1]{>{\raggedleft\let\newline\\\arraybackslash\hspace{0pt}}m{#1}}

\usepackage{siunitx}

\usepackage{enumitem}

\DeclareMathOperator{\area}{area}
\DeclareMathOperator{\closest}{closest}
\DeclareMathOperator{\dist}{dist}

\permission{}
\conferenceinfo{WWW 2016,}{April 11--15, 2016, Montr\'eal, Qu\'ebec, Canada.}
\copyrightetc{}
\crdata{}

\begin{document}

\title{The Death and Life of Great Italian Cities:\\A Mobile Phone Data Perspective}
\subtitle{}
%
%
%
%
%

\numberofauthors{6} 
%
\author{
%
%
\alignauthor
Marco De Nadai\\
       \affaddr{Fondazione Bruno Kessler}\\
       \affaddr{University of Trento}\\
       \affaddr{Trento, Italy}\\
       \email{denadai@fbk.eu}
\alignauthor Jacopo Staiano\\
       \affaddr{Sorbonne Universit\'{e}s}\\
       \affaddr{Paris, France}\\
       \email{jacopo.staiano@lip6.fr}
\alignauthor Roberto Larcher\\
       \affaddr{SKIL, Telecom Italia}\\
       \affaddr{Trento, Italy}\\
       \email{roberto.larcher@telecomitalia.it} 
\and  
\alignauthor Nicu Sebe\\
       \affaddr{University of Trento}\\
       \affaddr{Trento, Italy}\\
       \email{sebe@disi.unitn.it}
\alignauthor Daniele Quercia\\
       \affaddr{Bell Labs}\\
       \affaddr{Cambridge, United Kingdom}\\
       \email{quercia@cantab.net}
\alignauthor Bruno Lepri\\
       \affaddr{Fondazione Bruno Kessler}\\
       \affaddr{Trento, Italy}\\
       \email{lepri@fbk.eu}  
}

\date{31 January 2016}

\maketitle
\begin{abstract}
The Death and Life of Great American Cities was written in 1961 and is now one of the most influential book in city planning. In it, Jane Jacobs proposed four conditions that promote life in a city. However, these conditions have not been empirically tested until recently. This is mainly because it is hard to collect data about ``city life". 
The city of Seoul recently collected pedestrian activity through surveys at an unprecedented scale, with an effort spanning more than a decade, allowing researchers to conduct the first study successfully testing Jacobs's conditions.
In this paper, we identify a valuable alternative to the lengthy and costly collection of activity survey data: mobile phone data. We extract human activity from such data, collect land use and socio-demographic information from the Italian Census and Open Street Map, and test the four conditions in six Italian cities. Although these cities are very different from the places for which Jacobs's conditions were spelled out (i.e., great American cities) and from the places in which they were recently tested (i.e., the Asian city of Seoul), we find those conditions to be indeed associated with urban life in Italy as well.  Our methodology promises to have a great impact on urban studies, not least because, if replicated, it will make it possible to test Jacobs's theories at scale.

\end{abstract}




\section{Introduction}
By 2030, 9\% of the world population is expected to live in just $41$ \emph{mega-cities}, each of which will accommodate more than 10M inhabitants\footnote{\url{http://www.economist.com/node/21642053}}. 
Cities are  essential crucibles for innovation, tolerance, novelty, and economic prosperity~\cite{bettencourt2007growth, glaeser2011}, but they also face challenges in terms of, for example, traffic, pollution, poverty, and criminality~\cite{glaeser2011}. 

A fundamental research question that urban planners, sociologists and economists are investigating is what creates urban life. The urban sociologist Jane Jacobs in \emph{The Death and Life of Great American Cities}~\cite{jacobs1961death}, one of the most influential books in city planning, introduced the urban physical environment (the \emph{urban fabric}) as an essential factor for urban vitality~\cite{jacobs1961death}.
As the book title says, Jacobs dealt with the death and life of American cities. In its most basic form, her argument was that death was caused by elimination of pedestrian activity (e.g., by highway construction, large-scale development projects), and that life was created by the presence of pedestrians at all times of the day.

She argued that, to promote urban life in large cities, the physical environment should be characterized by \emph{diversity} at both the district and street level. Diversity, in turn,  requires four essential conditions: (i) \emph{mixed land uses}, that is, districts should serve more than two primary functions, and that would attract people who have different purposes; (ii) \emph{small blocks}, which promote contact opportunities among people; (iii) \emph{buildings diverse in terms of age and form}, which make it possible to mix high-rent and low-rent tenants; and (iv) sufficient \emph{dense concentration} of people and buildings.

Despite their importance, those conditions have not been empirically tested all together until recently, mainly because it is hard to collect data about which neighborhoods have full urban life and which have little of it. After an effort lasting more than a decade, the city of Seoul recently managed to collect pedestrian activity through surveys, and local researchers performed the first study testing Jacobs's conditions in the city~\cite{Sung01062015}. The researchers found her claims to hold in Seoul too: mixed use, old buildings, and density all contributed to urban life. 

To complement the lengthy and costly survey-based collection of activity data, this work put forward the use of mobile phone data. We extracted human activity measurements from mobile phone records in six Italian cities - Bologna, Florence, Milan, Palermo, Rome, and Turin. These cities are very different from the places in which Jacobs's conditions were spelled out (i.e., great American cities) and from the places in which they were recently tested (i.e., the Asian city of Seoul). Despite that, our results showed that the four conditions for promoting urban life do hold for Italian cities as well. To produce those results, we made three main contributions:
\begin{itemize}
\item Building upon previous works~\cite{jacobs1961death, Sung01062015}, we proposed a set of metrics for \emph{urban diversity} derived from open, aggregated, and anonymous map data, and a metric for \emph{urban vitality} derived from mobile phone activity (\Cref{sec:methodology});
\item We provided a comprehensive assessment of the predictive power of Jacobs's diversity conditions in the Italian context (\Cref{sec:results});
\item Finally, we identified our work's theoretical and practical implications, and its limitations (\Cref{sec:discussion}).
\end{itemize}

\section{Related Work}

Our work is best placed in an emerging interdisciplinary field called ``urban computing"~\cite{zheng2014}. This combines computer science approaches with more traditional fields like urban planning, urban economy, and urban sociology. 

The idea of testing urban theories using novel sources of data (e.g., social media, online images and videos, mobile phone data) has received increasing attention~\cite{quercia2014cscw,quercia2013www, hidalgo2014plos}. The urban sociologist Kevin Lynch showed that people living in an urban environment create their own personal ``mental map'' of the city based on features such as the areas they visit and the routes they use~\cite{lynch1960}. Hence, he hypothesized that, the more recognizable a city, the more navigable the city. To test Lynch's theory, Quercia \emph{et al.} built a web game that crowd-sourced Londoners' mental images of the city~\cite{quercia2013www}. They showed that areas suffering from social problems such as poor living conditions and crime are rarely present in residents' mental images.

Researchers also investigated which urban elements people use to visually judge a street to be safe, wealthy, and attractive using web crowdsourcing games~\cite{porzi2015multimedia,quercia2014cscw,hidalgo2014plos}, and they also studied how to identify walkable streets using  the social media data of Flickr and Foursquare (e.g., unsafe streets tended to be photographed during the day, while walkable streets were tagged with walkability-related keywords~\cite{quercia2015digital}).

The recent availability of large-scale data sets, such as those automatically collected by mobile phone networks, opens new possibilities for studying city dynamics at finer and unprecedented granularity~\cite{blondel2015}. Mobile phone data represents a highly valuable proxy for human mobility patterns~\cite{gonzalez2008,kung2014exploring,song2010science}. Such data was recently used to map functional uses~\cite{grauwin2015towards, lenormand2015comparing}, to identify places that play a major role in the life of citizens~\cite{isaacman2011identifying}, to compare cities based on their spatial similarities and differences~\cite{louail2014mobile}, and to predict socio-economic indicators~\cite{eagle2010, smith2014poverty}, including crime~\cite{bogomolovmoves, bogomolov2014, traunmeller2014}.


Researchers recently tested Jane Jacobs's observations about  crime~\cite{bogomolovmoves, bogomolov2014, traunmeller2014}. They verified her \emph{natural surveillance} theory. She posited that people diversity and presence of visitors in a neighborhood contribute to natural surveillance and, as such, discourage crime: as people are moving around an area, they effectively become ``eyes on the street" able to observe what is going on around them. 

To go beyond crime, we set out to conduct a comprehensive validation of Jacobs's four conditions for urban life. To do that, we gather several sets of data.


\vfill\eject
\section{Sets of data}
\label{sec:data}

\mbox{ } \\
\textbf{Mobile Phone Activity.} Every time a mobile phone communicates,  a radio base station delivers the communication through the network, and  a new call data record is created. This record reports the time of the communication, and the radio base station that handled it.
Different types of call records exist, recording various types of communication (e.g., incoming calls, Internet, outgoing SMS). 
We decided to focus on Internet activity, since it allows for the passive reconstruction of people's mobility: even in the absence of direct user activity (e.g., making/receiving a call, receiving/sending an SMS), mobile phones can be tracked since they are likely to be  connected to the Internet for background traffic and push notifications.

Call records have been provided by the Semantics and Knowledge Innovation Lab of Telecom Italia Mobile, which is the largest mobile operator in Italy with $34\%$ of the entire market share\footnote{\url{http://bit.ly/1LtNrFY}}.
Our data is aggregated every 60 minutes, comes from both TIM customers and roaming customers in the six cities, and covers the time ranging from February to October 2014.

\mbox{ } \\
\textbf{OpenStreetMap.} OpenStreetMap\footnote{\url{http://www.openstreetmap.com}} (OSM) is a global project that aims to collaboratively create a detailed map of the world. With more than 2.1 million contributors, it has become the most valuable and openly accessible source of information about the physical world. OSM data is composed by three data primitives: \emph{nodes}, \emph{ways} and \emph{relations}. \emph{Nodes} define points in space like Point Of Interests (which we call ``places''), \emph{ways} define roads and other linear features, and \emph{relations} reflect how those elements are related to each other. Since we focused on the geographic unit of a \emph{block} (i.e., the smallest area that is surrounded by street segments), we filtered out suburban elements: \emph{ways} with tag \texttt{highways} using Overpass API\footnote{\url{http://overpass-turbo.eu/}}, \texttt{footpaths}, \texttt{primary} roads, and \texttt{proposed} roads.

\mbox{ } \\
\textbf{Census Data.} The Italian National Institute for Statistics (ISTAT) provides, under open data licences, information gathered via the 2011 Italian Census\footnote{\url{http://www.istat.it/it/archivio/104317}}. We used  the census' variables related to people and  buildings (\Cref{tab:istat}), which were defined at various geographic levels.

\begin{table}[t!]
    \centering
    \small
    \begin{tabularx}{\columnwidth}{@{}lXl@{}}
        \midrule
        \textbf{ISTAT code} & \textbf{Description} \\ \toprule
        P1				& number of residents \\
        E2 				& number of used buildings \\
        E3 				& number of residential buildings \\
        E8-E16 		& number of buildings in 9 age bands in the range $[1919, 2011]$ \\
        E17-E20 		& number of buildings of 4 types based on their number of floors \\
        E21-E26 		& number of buildings of 6 types based on their internal apartments \\ 
        ADDETTI 		& number of employees \\ \bottomrule
    \end{tabularx}
    \caption{ISTAT open data variables under study. The last variable comes from the census of industry and services (\emph{censimento dell'industria e dei servizi}), while the rest come from the census of population and housing (\emph{censimento della popolazione e delle abitazioni}).}
    \label{tab:istat}
\end{table}


\mbox{ } \\
\textbf{Land Use.} Urban land-cover and land-use for Large Urban Zones (LUZs) is mapped by the  Urban ATLAS European project\footnote{\url{http://www.eea.europa.eu/data-and-maps/data/urban-atlas}}.
This project exploits satellite images to categorize the city in 20 classes (e.g., agricultural areas, continuous urban fabric) with a precision between 0.25 and 1 hectares, and accuracy above 80\%.
The dataset is built and validated for 2006, while its version updated to 2012 will be released at the end of 2015. However, for Turin and Rome, the new data is already available, and we used it. 

\mbox{ } \\
\textbf{Infrastructures.} In addition to the Urban ATLAS dataset, we used the ISTAT statistical atlas of infrastructures (\emph{Atlante statistico delle infrastrutture})\footnote{\url{http://www.istat.it/it/archivio/41899}}, which provides details on the logistic facilities (e.g., presence of railways) in Italy.

\mbox{ } \\
\textbf{Foursquare Data.} Created in 2008, Foursquare is the world leading location-based social network. It has attracted more than 30 million users\footnote{\url{http://bit.ly/1VNf4jQ}}.
Through a gamification system, users are encouraged to \emph{check-in} their geographical position into the social network and sharing the places they visit (\emph{venues}). 
Venues are places whose geographical information is enriched with semantic labels specifying details about them. 
Three hierarchical levels for place categories exist in Foursquare: an abstract level grouping \emph{venues} in macro categories (e.g., ``food"); a more specific one (e.g., ``restaurant"); and a third highly detailed level (e.g., ``Thai restaurant")\footnote{\url{https://developer.foursquare.com/categorytree}}. We extracted Foursquare  \emph{venues} from the public API\footnote{\url{https://api.foursquare.com}} for the categories  in~\Cref{tab:foursquareCategories}. 

\begin{table}[ht]
    \small
    \centering
    \begin{tabularx}{\columnwidth}{@{}lX@{}}
        \midrule
        \textbf{Group} & \textbf{Foursquare category} \\ \toprule

        NightLife 				& Brewery, Champagne Bar, Cocktail Bar, Dive Bar, Gay Bar, Hookah Bar, Lounge, Night Market, Night club, Other Nightlife, Pub, Strip Club, Whisky Bar, Wine bar, Nightlife Spot \\
        Art-night						& Laser Tag, Movie Theater, Music Venue, Performing Arts \\
        
        Services 				& Medical Center, Library, Government Building, Military Base, Post Office \\

        Eating-drinking				&	Food \\
        Org. activity		&	Comedy Club, Country Dance Club, Salsa Club, Club, Community Center, Cultural Center, Library, Social Club, Spiritual \\
        Outside						&	Park, Plaza, Pedestrian Plaza  \\
        Commercial					& Shop \& Service (excluding ATM, Construction, EV Charging, Gas Station / Garage, Newsstand) \\
         \bottomrule
    \end{tabularx}
    \caption{The Foursquare categories of the venues under study.}
    \label{tab:foursquareCategories}
\end{table}

\vfill\eject
\section{Death and Life of Cities}
In \emph{The Death and Life of Great American Cities}~\cite{jacobs1961death}, Jane Jacobs argued that ``death" in urban areas is caused by the elimination of pedestrian activity from urban streets. By contrast, urban life is created by a continual \emph{sidewalk ballet} consisting of a street filled with pedestrians at all times of the day.
In her own words, ``a well-used city street is apt to be a safe street and a deserted city street is apt to be unsafe"~\cite{jacobs1961death}.
This means that streets must be used fairly continuously during the day, both by residents and by strangers. These ``eyes on streets" generate a virtuous loop which, in turn, increases public safety, creates diverse face-to-face interactions, and contributes to  the local economy.

Jacobs spelled out four diversity requirements 
essential for the generation and maintenance of urban life. First, a district should serve more than one primary function, preferably more than two. This attracts people with diverse purposes who end up sharing common facilities. Moreover, when primary uses are  combined in a way that attracts  people at different times of the day, then that positively impacts the local economy.  For instance, a neighbourhood where people are present only during office hours is likely to provide only a few, if any, leisure facilities: it would be neither efficient nor attractive for residents.
Conversely, should a secondary function flourish by, e.g., the presence of theaters, museums or night-life places, residents would likely benefit. 

Second, Jacobs argued that street blocks must be short. She observed that large rectangular blocks are apt to thwart the effective mixture of use and people, resulting into paths that meet each other too infrequently. By contrast, smaller blocks provide more intersections, and thus slow down automobiles, which, she argued, are inclined to destroy urban vitality by discouraging the presence of pedestrian activity.

Third, in Jacobs's view, the buildings in a district should  be diverse in terms of age and form. This ensures diversity of economic activity. Diverse buildings make it possible to have a mixture of jobs (i.e., high-, medium- and moderate-income jobs) and a mixture of tenants (e.g., high-rent and low-rent tenants)~\cite{fainstein2000new, gordon2011does, jacobs1961death, king2013jane}. 
Her view came from observing large-scale projects in the New York of her time:  she saw that  large-scale buildings did not change over time and hardly adapted to the environment that surrounded them; by contrast, old buildings helped to cultivate new primary uses in the neighborhood (e.g., new companies were likely to start and initially grow in old and low-rent buildings).

The fourth and final condition is about dense concentration of people and buildings. The idea is that density fosters  ``lively" district that are able to attract people for different purposes.

Jacobs emphasized that \emph{all} four factors are necessary: density alone cannot create urban diversity, and mixed-land use would not flourish in areas with big blocks and of low density.

Further, in addition to those four conditions, Jacobs talked about `vacuums', which are patches of land dedicated to a single use (e.g., transport facilities, parks). Those elements might be good or bad. Small railway stations, bus stops and small parks may act as hubs for the pedestrian activity; but, at the same time,  if not well managed, a park could be used only at a certain time of the day, exposing it to deprivation and criminality.

\vfill\eject
\section{Methodology}
\label{sec:methodology}
Our approach leverages mobile Internet activity from call data records to extract proxies for \emph{urban vitality}, and web data from public entities (e.g., national census) and commercial ones (e.g., Foursquare) to extract \emph{urban diversity} as per Jacobs's four conditions. Our goal is to study the relationship between urban vitality and diversity. Next, we spell out the metrics and the regression models we used to meet that goal.

In the 2011 Italian census, the smallest administrative unit of cities is represented by the census sections (\emph{sezioni di censimento}), also called \emph{blocks}. These sections are delimited by street segments and are grouped with other nearby sections to form a census area (\emph{area di censimento}) called \emph{district}. 
The boundaries of a district are drawn on the basis of socio-economic conditions in a way that districts can be compared with each other in terms of, for example, population: districts have between 13,000 and 18,000 inhabitants\footnote{We exclude districts that are not densely populated (e.g. the national park in Rome, the volcano in Naples) These are marked by the Italian census with an identifier \texttt{ace} equal to zero.}. This is a loose constraint, however: city administrations may ignore it and define their own districts. Empirically, we verify that, in our data, a district covers an average area of 2.47 \si{\square\km} with an average population density of 10,000 people per \si{\square\km} (\Cref{tab:censusStats}). Jacobs did not define any strict criterion concerning district size: she simply proposed that the edge of an administrative district should not exceed 2.4 km, and that each district should have a minimum population of 50,000 people. Compared to that definition, our districts are similar in terms of area and smaller in terms of population. 

\begin{table}[t!]
    \centering
    \small
    \begin{tabularx}{\columnwidth}{@{}Xccc@{}}
        \midrule
        \textbf{City} & \textbf{\#Districts} & \textbf{Size (avg)} & \textbf{Population (avg)} \\ \toprule
        Bologna & 23 & 3.34 & 15,918
        \\
        Florence & 21 & 2.89 & 16,633
        \\
        Milan & 85 & 1.72 & 14,551
        \\
        Palermo & 43 & 2.01 & 15,075 
        \\
        Rome & 146 & 3.24 & 17,312 
        \\
        Turin & 56 & 2.00 & 15,543 
        \\
         \bottomrule
    \end{tabularx}
    \caption{Number, average size (in \si{\square\km}), and population of the districts in our six cities.}
    \label{tab:censusStats}
\end{table}

\subsection{Jacobs's metrics}
\label{sec:jacobs-metrics}
We extracted several variables to quantify the four conditions (\Cref{tab:variablesDistribution}), and we will detail them next. 

\begin{table}[ht!]
    \centering
    \small
    \begin{tabularx}{\columnwidth}{@{}lXr@{}} 
        \textbf{Land use} & \textbf{Distribution}  & \textbf{$\mu$}
        \\ \toprule
        (\ref{eq:LUM}) Land use mix				& \includegraphics[width=0.28\columnwidth]{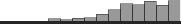} & 0.73 \\
        (\ref{eq:CSM}) Closeness to small parks (SPs) & \includegraphics[width=0.28\columnwidth]{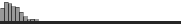} & \num{1d-3} \\
        (\ref{eq:RNR}) Residential \emph{vs.} Non-Res.		& \includegraphics[width=0.28\columnwidth]{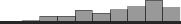} & 0.67 \\
        (\ref{eq:HT}) Housing types 	& \includegraphics[width=0.28\columnwidth]{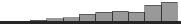} & 4.98 \\
(\ref{eq:C}) Commercial & \includegraphics[width=0.28\columnwidth]{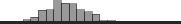} & 0.30
\\ 
(\ref{eq:N}) Nightlife & \includegraphics[width=0.28\columnwidth]{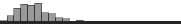} & 0.10
\\
(\ref{eq:ND}) Nightlife density  & \includegraphics[width=0.28\columnwidth]{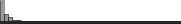} & \num{1d-5}
\\
(\ref{eq:D}) Daily & \includegraphics[width=0.28\columnwidth]{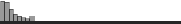} & 0.02 \\
(\ref{eq:TP}) $3^{rd}$Places  & \includegraphics[width=0.28\columnwidth]{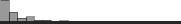} & \num{1d-4}
\\ 
\\
\textbf{Small blocks}
\\
\midrule
(\ref{eq:BA}) Block area & \includegraphics[width=0.28\columnwidth]{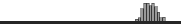} & 9.61
\\ 
(\ref{eq:ID}) Intersections density & \includegraphics[width=0.28\columnwidth]{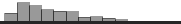} & \num{1d-4}
\\
(\ref{eq:phi}) Anisotropicity & \includegraphics[width=0.28\columnwidth]{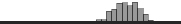} & 0.38
\\
\\
\textbf{Aged buildings}
\\
\midrule
(\ref{eq:avg}) $\overline{\text{building age}}$ & \includegraphics[width=0.28\columnwidth]{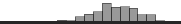} & 50.34
\\ 
(\ref{eq:age}) $\sigma_{\text{building age}}$ & \includegraphics[width=0.28\columnwidth]{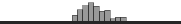} & 12.96
\\ 
(\ref{eq:EC}) $\overline{\text{Employees per company}}$ & \includegraphics[width=0.28\columnwidth]{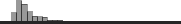} & 6.98
\\
\\
\textbf{Concentration}
\\
\midrule
(\ref{eq:PD}) Population density & \includegraphics[width=0.28\columnwidth]{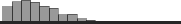} & 0.01
\\
(\ref{eq:ED}) Employment density & \includegraphics[width=0.28\columnwidth]{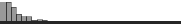} & \num{5d-3}
\\
(\ref{eq:PDED}) $\frac{\text{population density}}{\text{employee density}}$ & \includegraphics[width=0.28\columnwidth]{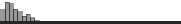} & 3.72
\\
(\ref{eq:IN}) $\frac{|\text{internal apartments}|}{|\text{buildings}|}$ & \includegraphics[width=0.28\columnwidth]{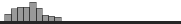} & 15.31
\\
(\ref{eq:DDP}) Density of daily places & \includegraphics[width=0.28\columnwidth]{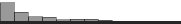} & \num{5d-3}
\\
(\ref{eq:DNDP}) Density non-daily places & \includegraphics[width=0.28\columnwidth]{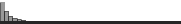} & \num{3d-3}
\\
\\
\textbf{Vacuums}
\\
\midrule
(\ref{eq:CLP}) Closeness to large parks & \includegraphics[width=0.28\columnwidth]{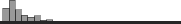} & \num{3d-3}
\\ 
(\ref{eq:CR}) Closeness to railways & \includegraphics[width=0.28\columnwidth]{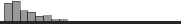} & \num{1d-3}
\\ 
(\ref{eq:CH}) Closeness to highways & \includegraphics[width=0.28\columnwidth]{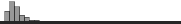} & \num{6d-4}
\\ 
(\ref{eq:CW}) Closeness to water & \includegraphics[width=0.28\columnwidth]{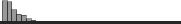} & \num{1d-3}
\\ 
\\
\textbf{Mobile phone activity}
\\
\midrule
(\ref{eq:activity}) Activity density & \includegraphics[width=0.28\columnwidth]{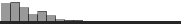} & 3.84
\\ 
    \end{tabularx}
    \caption{Urban diversity metrics plus mobile Internet activity.
    Most variables are not normally distributed. Density measures are computed with surface area in \si{\square\meter}, and closeness measures are the inverse of the distance (\si{1/\meter}).}
    \label{tab:variablesDistribution}
\end{table}

\subsubsection{Land use}
\label{sec:mixOfLandUses}
The first of Jacobs's four conditions is to have mixed primary uses in a district~\cite{jacobs1961death}. Primary categories include residential buildings, offices, industrial facilities, entertainment places, education facilities, recreation facilities, museums, libraries, and galleries. 



We computed Land Use Mix (LUM)~\cite{cervero1989land} in district $i$ as:
\begin{equation}
\text{LUM}_i = - \sum_{j=1}^n \frac{P_{i,j} \log(P_{i,j})}{\log(n)}
\label{eq:LUM}
\end{equation}
where $P_{i,j}$ is the percentage of square footage with land use $j$ in district $i$, and $n$ is the number of possible land uses (in our case, $n=3$). If district $i$'s land is dedicated to one use only, then $LUM_i$ is zero; instead, if the land is used equally in all $n$ ways, then $LUM_i$ is one. The higher $LUM_i$, the more mixed $i$'s land use. We defined the three land uses following the recommendations in~\cite{manaugh2013mixed}: the first land use is ``residential"; the second includes the categories ``commercial", ``institutional and governmental", and ``resource and industrial"; and the third includes ``park and recreational" and ``water".

Since well-managed parks might well function as hubs for pedestrian activity,  we defined the average distance from the nearest small park (area smaller than $1$ \si{\square\km}) for each district $i$:
\begin{equation}
\text{Closeness to SM}_i = (\frac{1}{|B_i|} \sum_{j \in B_i} \dist(j, \closest(j, SM)))^{-1}
\label{eq:CSM}
\end{equation}
where $B_i$ is the set of blocks in district $i$, $\closest(j, Y)$ is a function that finds the geographically closest element in set $Y$ from block $j$'s centroid, $SM$ is the set of small parks, and $\dist(a,b)$ is the geographic distance between two elements' centroids $a$ and $b$.


We also computed the Residential/Non-Residential (RNR) balance in district $i$ as:
\begin{equation}
\text{RNR}_i = 1- | \frac{Res_i - NonRes_i}{Res_i + NonRes_i}|
\label{eq:RNR}
\end{equation}
where $Res_i$ is the area occupied by residential buildings in district $i$, and $NonRes_i$ is the area occupied by non-residential ones. The higher $RNR_i$, the more balanced the district in terms of residential \emph{vs.} non-residential uses.

To go beyond horizontal land use and look at vertical development, we computed the average number of floors per building in district $i$ and called it `housing types' (as~\cite{Sung01062015} did):
\begin{equation}
\text{Housing types}_i = \frac{\sum_{c } h_{c,i}\cdot z_c}{\sum_{c } h_{c,i}}
\label{eq:HT}
\end{equation}
where $h_{c,i}$ is the number of buildings that are in height category $c$ in district $i$, and $z_c$ is the number of floors corresponding to height  category $c$. The sums were repeated over all height categories (i.e., over the categories \texttt{E17-E20} in \Cref{tab:istat}).


The previous definitions have characterized spatial use in terms of land use and building use. However, activities are important too. Jacobs argued for mixing primary uses so that people are on the street at different times of the day. To characterize spatial use in terms of activities, upon Foursquare data, we determined whether each place is used daily (e.g., convenience stores, restaurants, sport facilities) or not, and whether it is used at night or daylight~\cite{Sung01062015}. Based on that, we defined:
\begin{equation}
\text{Commercial}_i = \frac{|\text{non daily-use places}_i|}{|\text{places}_i|}
\label{eq:C}
\end{equation}
\begin{equation}
\text{Nightlife}_i = \frac{|\text{nightlife places}_i|}{|\text{places}_i|}
\label{eq:N}
\end{equation}
\begin{equation}
\text{Nightlife density}_i = \frac{|\text{nightlife places}_i|}{\area_i}
\label{eq:ND}
\end{equation}
\begin{equation}
\text{Daily}_i = (\frac{1}{|B_i|} \sum_{j \in B_i} \dist(j, \closest(j, D)))^{-1}
\label{eq:D}
\end{equation}
where $D$ is the set of places that are used daily, and $B_i$ is, again, the set of street blocks in district $i$.


Also, not all activities are equal. There are activities that are more `social' than others. To capture that, we resorted to the concept of \emph{third places}. These are defined by Oldeburg~\cite{oldenburg1989great} as the ``great, good places" that foster community and communication among people outside home (the first place) and work (the second place): ``they are places where people gather primarily to enjoy each others' company". Third Places function as unique public spaces for social interaction, providing a context for sociability, spontaneity, community building and emotional expressiveness~\cite{oldenburg1982third}. Therefore, we  computed:
\begin{equation}
3^{rd}places_i = \frac{|3^{rd} \text{places}_i|}{|\text{places}_i|}
\label{eq:TP}
\end{equation}
We determined whether a place is a third place or not by following the 4-category classification proposed by Jeffres \emph{et al.}~\cite{jeffres2009impact}. Third places fall into these four categories: \emph{eating, drinking and talking} (e.g., coffee shops, bars, pubs, restaurants, and cafes);  \emph{organized activities} contributing to social capital~\cite{putnam2001bowling} (e.g., places of worship, clubs, organizations, community centers, and senior centers); \emph{outdoor} (e.g., plazas and parks); and \emph{commercial venues} (e.g., stores, malls, shopping centers, markets, beauty salons, and barber shops).

\subsubsection{Small blocks}
Jacobs listed the presence of small blocks as the second necessary condition for diversity. Small blocks are believed to support stationary activities and provide opportunities for short-term and low-intensity contacts, easing interactions with other people in a relaxed and relatively undemanding way. Specifically, she stated that ``lowly, unpurposeful and random as they may appear, sidewalk contacts are the small change from which a city's wealth of public life may grow''. She criticized super-blocks and rectangular blocks, which constrain urban mobility with high travel distances and limited opportunities of cross-use.

The easiest way to identify small blocks is to compute the average block area among the set $B_i$ of blocks in district $i$:
\begin{equation}
\overline{\text{Block area}}_i = \frac{1}{|B_i|} \sum_{j \in B_i} \area_j
\label{eq:BA}
\end{equation}

Since a district with high intersections density is likely to contribute to random contacts, we also computed:
\begin{equation}
\text{Intersection density}_i = \frac{|\text{intersections}_i|}{\area_i}
\label{eq:ID}
\end{equation}

Finally, since block size is distributed as a power law $P(A) \sim \frac{1}{A^r}$ with $r \sim 2$, we characterized a district $i$ by its average shape anisotropicity~\cite{louf2014typology} of the blocks $B_i$ within it:
\begin{equation}
\label{eq:phi}
\text{District anisotropicity}_i = \frac{1}{|B_i|} \sum_{j \in B_i} \Phi_j
\end{equation}
where $\Phi_j$ is the ratio between the area of the block $j$ and the area of its circumscribed circle $\mathcal{C}_j$:
\begin{equation}
\Phi_j=\frac{\area_j}{\area_{\mathcal{C}_j}}
\end{equation}
The quantity $\Phi_j$ is always smaller than one, and the larger its value, the less anisotropic block $j$, the more opportunities for contacts the block creates.

All those metrics were computed using the district \emph{net} area $area_i$, which excludes unpopulated patches such as rivers and natural parks.


\subsubsection{Aged buildings}
Jacobs stressed the importance of having old buildings in a district. If a district has only new buildings, then it would have only enterprises that  can support the high costs of new constructions. If it has old buildings too, instead, it would be able to  incubate new small enterprises that cannot afford high rents, and that will benefit the local economy in the long run: ``If the incubation is successful enough, the yield of the building can, and often does, rise"~\cite{jacobs1961death}.


As a first measure, we computed the average building age in district $i$:
\begin{equation}
\overline{\text{Building age}_i}  = \frac{\sum_{b} |\text{building}_{b,i}| \cdot \text{old}_b}{\sum_{b} |\text{building}_{b,i}|}
\label{eq:avg}
\end{equation}
This measures how old, on average, a building in district $i$ is. Since age is expressed in age bands in the Italian census (the first band is \texttt{E8} in Table~\ref{tab:istat}, and the last is \texttt{E16}), we needed to compute the temporal length of each band. More specifically, for each band $b$ in the range $[\text{start}_b,\text{end}_b ]$, we computed:
\begin{equation}
\text{old}_b = \frac{(\text{last}-\text{start}_b) + (\text{last}-\text{end}_b)}{2}
\end{equation}
where last is the year of the latest built building in the whole census data. Then, in equation~(\ref{eq:avg}), we weighted age band $b$'s temporal length ($\text{old}_b$) by the number of buildings in that band ($|\text{building}_{b,i}|$), and that gave us the average age of the district's buildings. 

We also computed the corresponding standard deviation as:
\begin{equation}
\sigma_{\text{building age}_i}  = \frac{\sum_{b} |\text{building}_{b,i}|^2}{\left (\sum_{b} |\text{building}_{b,i}| \right)^2} \cdot \sigma^2
\label{eq:age}
\end{equation}
where $\sigma^2$ is the standard deviation of the number of buildings in each band. The higher the standard deviation, the higher the district's  heterogeneity in terms of building age.


Finally, we computed the average number of employees per company in district $i$:
\begin{equation}
\overline{\text{Employees per company}_i}= \frac{1}{|F_i|} \sum_{j \in F_i} \text{employees}(j)
\label{eq:EC}
\end{equation}
where $F_i$ is the set of companies in district $i$.

\subsubsection{Concentration}
Jacobs's fourth and final condition is about having concentration of (both residential and non-residential) buildings and of people. 
Similarly to Sung \emph{et al.}~\cite{Sung01062015}, we computed two sets of concentration measures: one for people, and the other for buildings. 

First, population and employment density measures were calculated dividing the number of people (employees) by the district's net area.

\begin{equation}
\overline{\text{Population density}_i}= \frac{|\text{Population}_i|}{\area_i}
\label{eq:PD}
\end{equation}
\begin{equation}
\overline{\text{Employment density}_i}= \frac{|\text{Employed people}_i|}{\area_i}
\label{eq:ED}
\end{equation}

We then computed  the interaction term between those two measures as the ratio between population density and employment density:
\begin{equation}
\frac{\text{Population density}_i}{\text{Employment  density}_i}
\label{eq:PDED}
\end{equation}
The higher it is, the more residents (as opposed to employees) in the district. Different values correspond to districts with different social textures.

To add to those people-based concentration measures one building-based measure, we computed:
\begin{equation}
\text{internal}_i= \frac{|\text{internal apartments}_i|}{|\text{buildings}_i|}
\label{eq:IN}
\end{equation}
which is the average number of apartments per building.

Finally, as we have done previously, to go beyond people and buildings and look at activities, we computed:

\begin{equation}
\text{Density of daily places}_i= \frac{|\text{daily-use places}_i|}{\area_i}
\label{eq:DDP}
\end{equation}

\begin{equation}
\text{Density of non-daily places}_i= \frac{|\text{non-daily-use places}_i|}{\area_i}
\label{eq:DNDP}
\end{equation}

These two quantities are not totally anti-correlated to each other since not all places can be classified as being fully daily \emph{vs.} non-daily.



\subsubsection{Vacuums}
Border vacuums are places that act as physical obstacles to pedestrian activity.  For instance, parks can be a hub of pedestrian activity, if efficiently managed~\cite{jacobs1961death}, but they could also be deplorable places in which criminality flourishes (especially at night). In a similar way, the proximity to expressways may discourage pedestrian activity or may effectively connect different parts of the city. This is what Jacobs called ``the curse of border vacuums". 

\begin{figure*}[ht!]
\centering
\includegraphics[width=\textwidth]{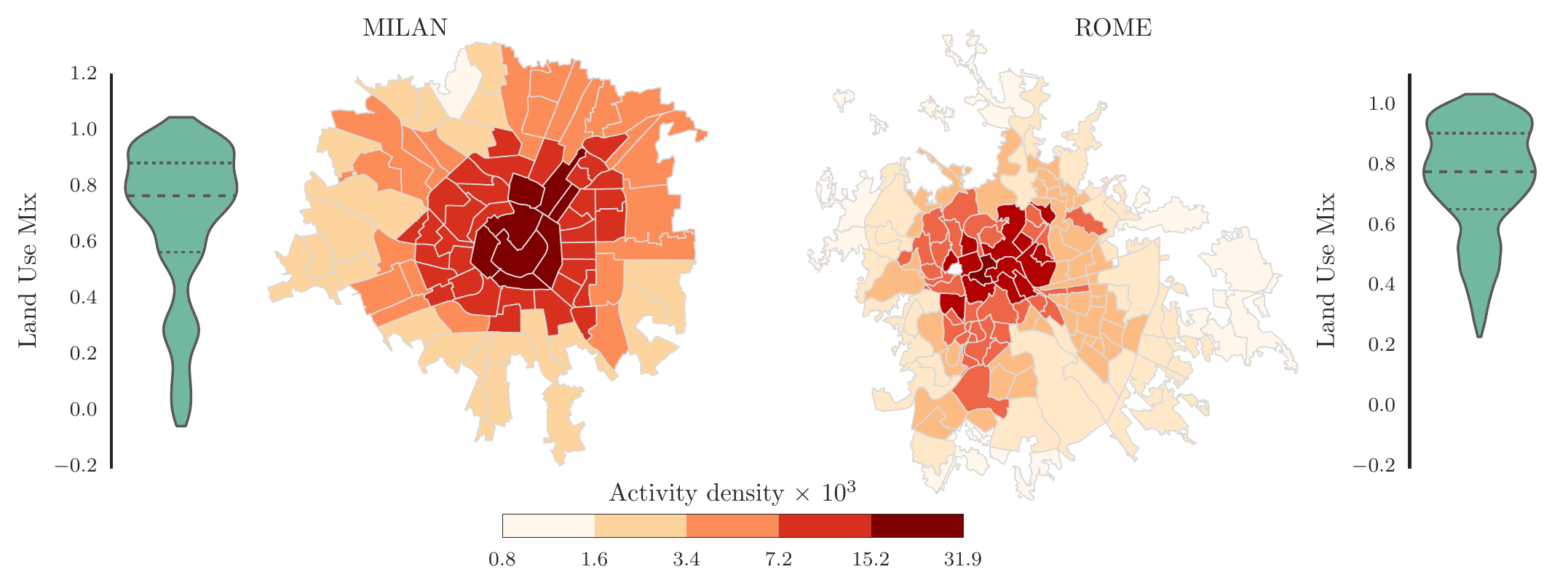}
\vspace{-1.5em}
\caption{District activity density in Milan (left) and Rome (right), and their corresponding values of mixed land use (LUM). The colors are in logarithmic scale.}
\label{fig:landusemixSignature}
\end{figure*}

Therefore, we needed to identify large areas with single use. From OSM, we took parks\footnote{We took areas identified by \texttt{leisure=park}, filtering out the parks with \texttt{barrier}.}, and, from  the Urban ATLAS, we took railway areas, fast transit zones, rivers, lakes, and highways. That extraction allowed us to build the sets of large parks $LP$, railways $R$, highways $H$ and water areas $W$. 

To verify the impact of a type of vacuum area, say, that of large parks on a district, we calculated the average distance between a district's block (i.e., smallest area surrounded by street segments) and its closest large park:
\begin{equation}
\text{Closeness to LP}_i = (\frac{1}{|B_i|} \sum_{j \in B_i} \dist(j, \closest(j, LP)))^{-1}
\label{eq:CLP}
\end{equation}
where $\dist(j, \closest(j, LP)$ is the distance between block $j$ and its closest large park. The sum of distances is done over all blocks in district $i$ ($B_i$ is indeed the set of district $i$'s blocks). 

The next type of vacuum area is that of railways. Our dataset did not differentiate between rails and stations. However, in terms of pedestrian activity, the two differ: often, rails act as obstacles of activity, while stations act as hubs. To incorporate that distinction, we excluded the stations (i.e., 
the 600-meter buffer areas around them) from the set $R$ of railways, and computed:
\begin{equation}
\text{Closeness to R}_i = (\frac{1}{|B_i|} \sum_{j \in B_i} \dist(j, \closest(j, R)))^{-1}
\label{eq:CR}
\end{equation}

In a similar way, we computed the  average distance between a district's block and its closest highway, and its closest water area:

\begin{equation}
\text{Closeness to H}_i = (\frac{1}{|B_i|} \sum_{j \in B_i} \dist(j, \closest(j, H)))^{-1}
\label{eq:CH}
\end{equation}

\begin{equation}
\text{Closeness to A}_i = (\frac{1}{|B_i|} \sum_{j \in B_i} \dist(j, \closest(j, A)))^{-1}
\label{eq:CW}
\end{equation}
where $H$ and $A$ are the sets of highways and of water areas.

\subsection{Activity density}
\label{sec:activity-density}

As it has been extensively done in previous work, we used mobile Internet activity as a proxy for \emph{urban vitality}.


To roughly estimate the number of Internet connections that fell into each district, we needed to represent the urban space somehow. We chose to represent it  as a set of 2-dimensional, non-overlapping, and non-convex polygons. These polygons come from a Voronoi tessellation based on the radio stations' positions. 

Then, to estimate the number of Internet call records $S_{i}(t)$ in each district $i$ at time $t$, we counted the number of Internet connections over all polygons $v$'s that fell into district $i$: 
\begin{equation}
	S_{i}(t) = \sum_{v } R_{v}(t) \frac{A_{v \cap i}}{A_v - A_{v \cap W}}
\label{eq:activity} 
\end{equation}

where $v$ is a polygon, and $R_{v}(t)$ is the number of Internet connections in $v$ at time $t$. The count of Internet connections is weighted by $\frac{A_{v \cap i}}{A_v - A_{v \cap W}}$, which is the proportion  $\frac{A_{v \cap i}}{A_v }$ of $v$'s area that falls into district $i$ ($A_{v \cap i}$ is $v$'s area that falls into district $i$, and $A_v$ is  $v$'s total area). From $v$'s total area we removed sea areas denoted by $W$ (i.e., we removed $A_{v \cap W}$). 

Then, having $S_{i}(t)$, we computed a district's \emph{activity density} as the average number of Internet connections throughout a typical business day, divided by the district's area. The results for Milan and Rome are shown in \Cref{fig:landusemixSignature}. The normalization of surface area makes it possible to compare the activities of districts of different sizes.



\subsection{The regression model}
We used an Ordinary Least Squares (OLS) model to evaluate the relationship between each district's structural diversity (Section~\ref{sec:jacobs-metrics}) and the district's activity density (Section~\ref{sec:activity-density}).

Since most of the regression variables were skewed, we transformed them. More specifically, we log-transformed  activity density using the natural logarithm, and transformed the structural diversity metrics using the Box-Cox method~\cite{box1964analysis}. To avoid over-fitting, we did split the data into training set (75\%) and test set (25\%), and repeated our measurements 1000 times using a shuffle split cross-validation.


We created six linear models. First, we created five models, each of which had \emph{one} of the five sets of Jacobs's metrics as independent variables (i.e., we separately analyzed land use, small blocks, aged buildings, concentration, and vacuums). Those models always had activity density as  dependent variable. The resulting coefficients are shown in the first five columns of \Cref{tab:regression}. 
We then had a combined model by selecting the independent variables through 
both recursive feature elimination and stability selection~\cite{meinshausen2010stability}. The resulting selected variables and corresponding coefficients are shown in the last column of \Cref{tab:regression}.


\section{Results}
\label{sec:results}

\subsection{Land use}
 Mixed land use matters only in cities in which functional uses were historically separated like in the case of Milan. By looking at the first two predictors of land use in Table~\ref{tab:regression}, one learns that mixed land use does not contribute neither positively nor negatively. That is because not all cities are equal: land use in Rome is mixed (Figure~\ref{fig:landusemixSignature}), while Milan is separated in functional areas. Consequently, in Milan, vitality is experienced only in the mixed districts. 
 However, the highest beta coefficient concerning land use is found to be related to the presence of third places ($\beta=0.3972$). Daily errands are important, but third places are more so (with a Spearman's correlation of $0.8337$ with activity density). These places are public places in which people can hang out for good company and lively conversations, putting aside the concerns of home and work (their first and second places)~\cite{oldenburg1989great}. Examples of such places include pubs, coffee shops, and taverns.

\begin{table*}[ht!]
\small
\centering
\ra{1.2}
\begin{tabular}{@{}rllllll@{}}
\toprule
&
Land use & Small blocks & Aged buildings & Concentration & Vacuums & Combined
\\
\midrule
$Adj R^2$ & 0.70 & 0.63 & 0.40 & 0.73 & 0.19 & 0.77
\\ \hline
Land use mix (\ref{eq:LUM}) & 0.0236
\\
Closeness to small parks (SPs) (\ref{eq:CSM}) & 0.0409 &&&&& 
\\
Residential \emph{vs.} Non-Res. (RNR) (\ref{eq:RNR})& -0.0180
\\
Housing types (\ref{eq:HT})& 0.1959*** &&&&& 0.1854***
\\ 
Commercial (\ref{eq:C}) & 0.0138
\\ 
Night-life (\ref{eq:N})& 0.0110
\\
Night-life density (\ref{eq:ND})& -0.1107*  
\\
Daily (\ref{eq:D}) & 0.0440
\\ 
$3^{rd}$Places (\ref{eq:TP}) & 0.3972*** 
\\
Block area (\ref{eq:BA})&& 0.0615
\\ 
Intersections density (\ref{eq:ID})&& 0.7380***  &&&& 0.1914***
\\
Anisotropicity (\ref{eq:phi})&&  0.0462
\\
$\overline{\text{building age}}$ (\ref{eq:avg})&&& 0.3689*** &&& 
\\ 
$\sigma_{\text{building age}}$ (\ref{eq:age})&&& 0.2264*** &&& 
\\ 
$\overline{\text{Employees per company}}$ (\ref{eq:EC})&&& -0.0778
\\
Population density (\ref{eq:PD})&&&& -0.1480
\\
Employment density (\ref{eq:ED})&&&&  0.7401*** && 0.4344***
\\
$\frac{\text{population density}}{\text{employee density}}$ (\ref{eq:PDED})&&&& 0.3020*
\\
$\frac{|\text{internal apartments}|}{|\text{buildings}|}$ (\ref{eq:IN})&&&& 0.0220  
\\
Density of daily places (\ref{eq:DDP})&&&& 0.2046**
\\
Density non-daily places (\ref{eq:DNDP})&&&& 0.1061
\\
Closeness large parks (\ref{eq:CLP})&&&&& 0.1384** 
\\ 
Closeness railways (\ref{eq:CR})&&&&&  0.1745***
\\ 
Closeness highways (\ref{eq:CH})&&&&&  -0.3457*** & -0.1018***
\\ 
Closeness water (\ref{eq:CW})&&&&& 0.0991
\\ 
\specialcell{$3^{rd}$Places $\times$ Closeness to highways} &&&&&&  0.0681** 
\\
\specialcell{Closeness to SPs $\times$ Closeness to highways} &&&&&&  -0.0788** 
\\
\bottomrule
\end{tabular}
\caption{Linear regression models that predict district activity density. Each column is a different model. For each model's column, the table reports the model's predictive power (Adjusted $\mathbf{R^2}$ in the first row) and the  $\mathbf{\beta}$ coefficients.
Blank cells correspond to variables that are absent from the model. As for statistical significance, we use the following notation: *$\mathbf{p<0.05}$, **$\mathbf{p<0.01}$, ***$\mathbf{p<0.001}$.}.
\label{tab:regression}
\end{table*}

\subsection{Small blocks}
Continental European cities do not have the super-blocks typical of American cities. This is especially true in Italy, so it comes at no surprise that block size does not greatly matter ($\beta = 0.0615$ in model ``small blocks'' of Table~\ref{tab:regression}). By contrast, the density of intersections does matter: vibrant urban areas are those with dense streets, which, in fact, slow down cars and make it easier for pedestrian to cross~\cite{calthorpe93}, creating what  Jane Jacobs called the ``sidewalk ballet''. This `ballet' goes beyond simple pedestrian activity. It is about  informal contacts and  public trust (e.g., children playing on the street, adults other than their parents paying attention to them, leaving home keys with shopkeepers). To increase the potential for such types of interactions, it is essential to be able to look into each others' eyes, and only small streets could foster that. 


\subsection{Aged buildings}
In large American cities, Jacobs observed that neighborhoods with aged buildings tended to have a diversified local economy. By mingling building of varying age and conditions (i.e., mixing expensive building with ``a good lot of plain, ordinary, low-value old buildings,  rundown old buildings''), a neighborhood attracts not only standardized and high-earning enterprises but also ordinary and innovative ones. Aged buildings were a rarity in American cities. Back in her days, Jacobs noted: ``in Miami Beach, where novelty is the sovereign remedy, hotels ten years old are considered aged and are passed up because others are newer''~\cite{jacobs1961death}. Decisions about which buildings get to stay and which get to go depend on culture: they are influenced by the relationship that a culture has with time and place. In a neighborhood, the patina of time may be \emph{retained}, \emph{imitated}, or \emph{removed}, as Kevin Lynch puts it~\cite{lynch-place-time}. In Western Europe, the idea of preservation appeared about 1500 and, since then, retention is by far preferred over removal. As a result, Italian districts are defined by age. 
Central areas are the ancient ones and their surrounding areas followed over time - each district ended up having its own era. Consequently, in the Italian context, mixing buildings of different eras is not as important as (or, rather, as possible as) it is in the American context.



\subsection{Concentration}
The last of Jacobs's conditions is about dense concentration of people and buildings. In our cities, the most informative purpose is that of office work (highest beta coefficient for the column named ``Concentration'' of Table~\ref{tab:regression}). This is reasonable as the previous three conditions (embedded in the regression) speak to the vitality contributed by \emph{residents}, and this fourth condition adds to it the vitality contributed by \emph{dwellers} who likely work in the district.


\subsection{Border vacuums}
Finally, looking at the ``vacuums" model in \Cref{tab:regression}, we observe little effect of border vacuums. Surprisingly, railways or rivers - which might hamper pedestrian activity at times - seem, instead, to be effectively integrated in the social fabric of active districts. However, as one expects, highways are not: in general, being close to them ends up being detrimental ($\beta = -0.3457$ in the ``vacuums" column and $\beta = -0.1018$ in the ``combined" column). 
\mbox{ } \\

\Cref{fig:alluvialPlot} shows to which extent (in terms of $\beta$ coefficients) each of Jacobs's four conditions explains district activity.

\subsection{Summary}

\begin{figure}[ht!]		
	\centering		
	\includegraphics[width=\columnwidth]{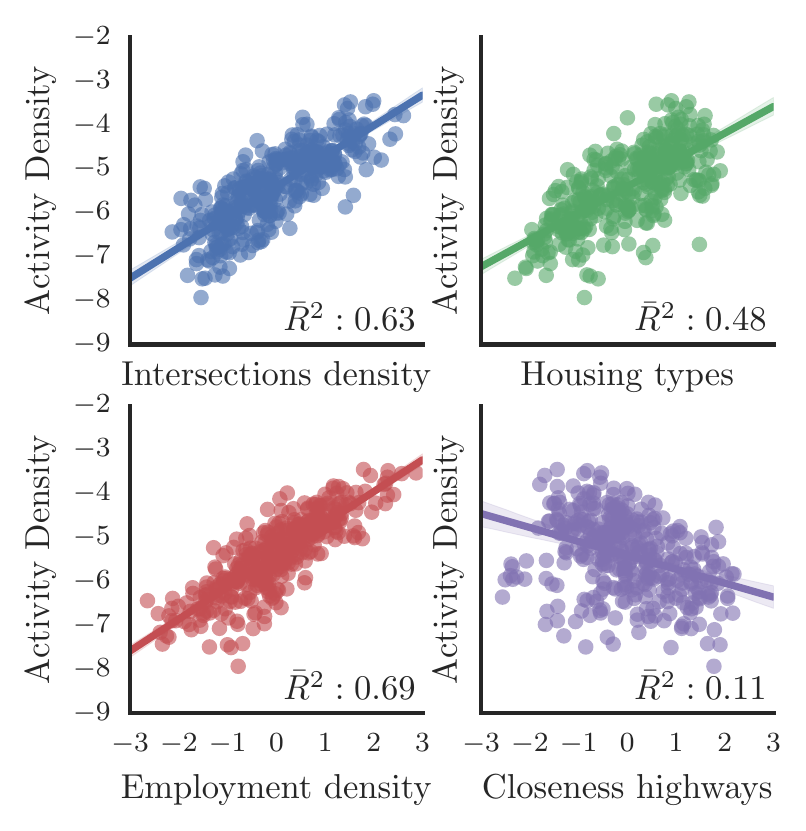}
	\vspace{-1.5em}
	\caption{Plots of district activity as a function of intersection density, housing types, employment density, and closeness to highways. A district's activity is strongly associated with the district's static structural features.}		
	\label{fig:lfit}		
\end{figure}

\begin{figure}[ht!]
\centering
\includegraphics[width=\columnwidth]{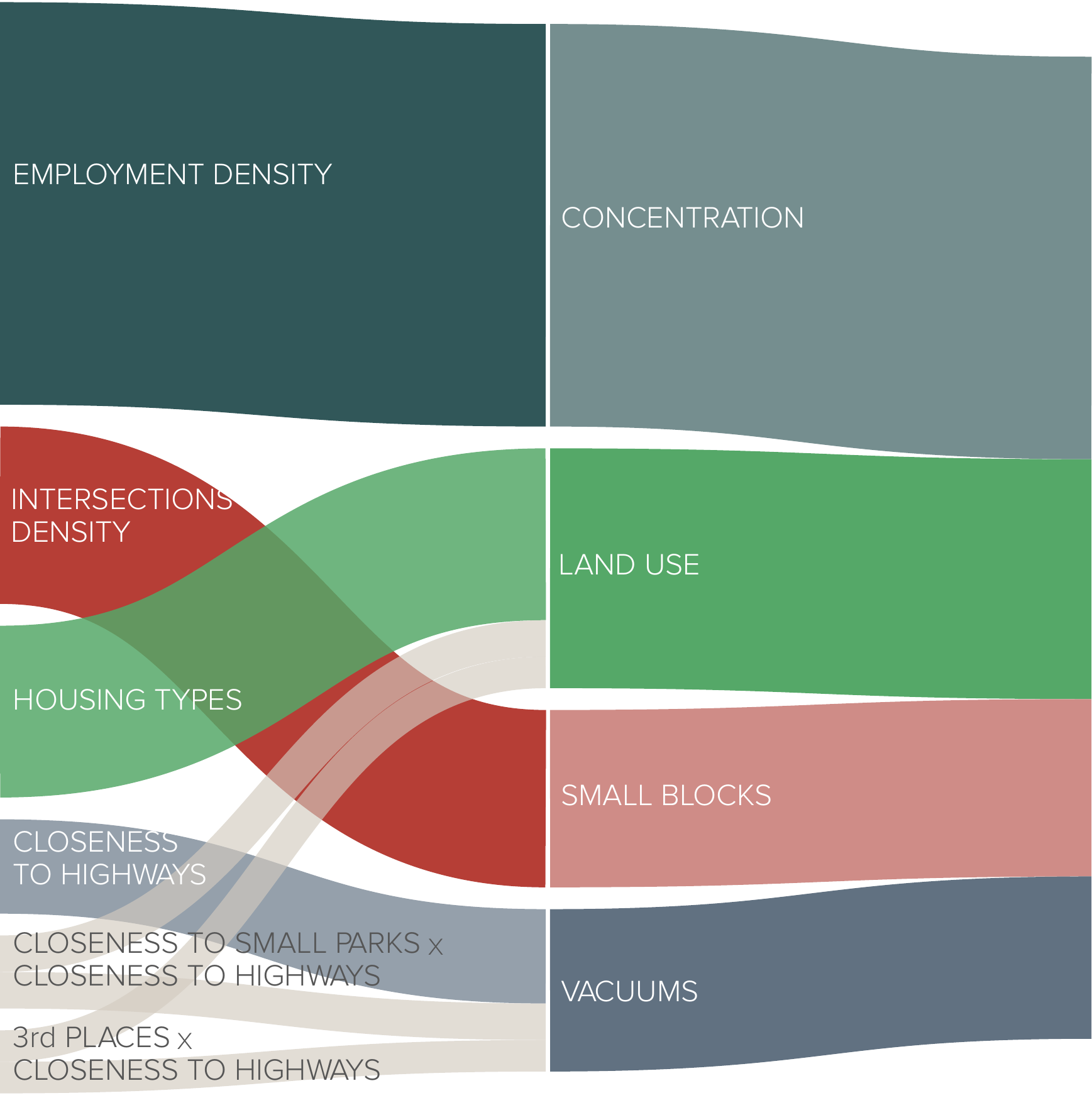}
\caption{The predictive power of  Jacobs's features. The bar size is proportional to the absolute $\beta$ value in the linear regression model.  The first column shows individual features, and the second shows grouped features. The three grouped features - concentration, land use, and small blocks - plus vacuum areas explain 77\% of the variability of activity density.}
\label{fig:alluvialPlot}
\end{figure}

Taken together, our results suggest that Jacobs's four conditions for maintaining a vital urban life hold for Italian cities as well (see \Cref{fig:alluvialPlot}). From \Cref{fig:alluvialPlot} and the last column of Table~\ref{tab:regression}, we see that as much as 77\% of the variability of district activity is explained by simple structural and static features. As Figure~\ref{fig:lfit} details, even individual features are strongly associated with activity. 

\begin{figure}[ht!]
\centering
\includegraphics[width=\columnwidth]{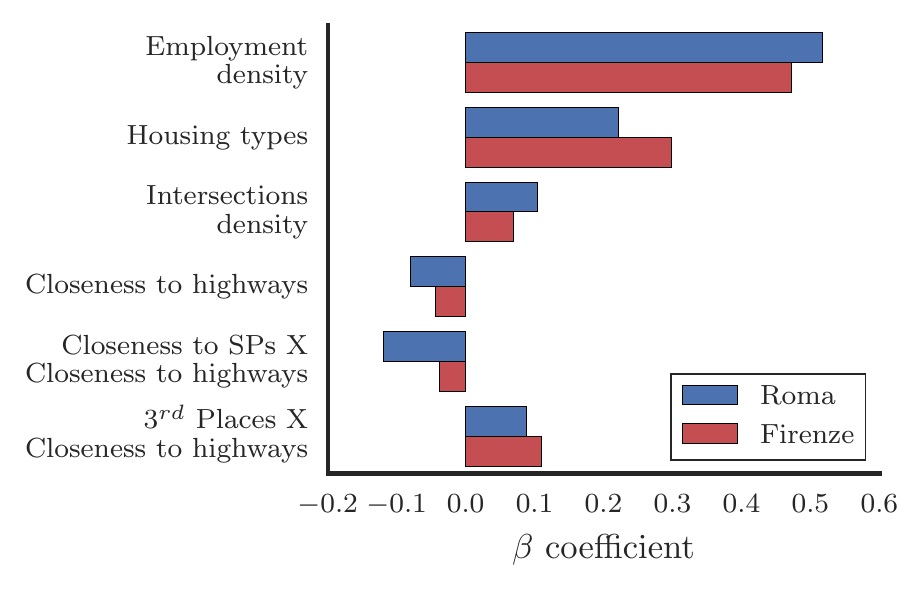}
\vspace{-1.5em}
\caption{The $\beta$ coefficients of two city-specific linear regression models for Roma and Firenze.}
\label{fig:barplot}
\end{figure}

Also, the extent to which the different features matter does not dramatically change across cities. To partly verify that, we took the largest and smallest cities: Rome and Firenze. As \Cref{fig:barplot} suggests, all the main features mattered to a very similar extent. 

Based on our findings, to paraphrase the four conditions in the Italian context, we might say that:

\mbox{ } \\
\fbox{\begin{minipage}{25em}
\emph{Active Italian districts have dense concentrations of office workers, third places at walking distance, small streets, and historical buildings.}
\end{minipage} }

\mbox{ } \\
\mbox{ } \\

\section{Discussion}
\label{sec:discussion}
After having operationalized Jane Jacobs's four conditions, we now discuss our work's practical and theoretical  implications, and the questions it left open.

\mbox{ } \\ 
\textbf{Practical Implications.} We foresee several practical implications of our work:

\begin{description}
\item \emph{City dashboards.} Part of our work could inform the selection of which features should go into a city dashboard. We have shown that a variety of structural features of the built environment are closely linked to district activity. Given the importance of those features and the ease with which they could be computed upon web and official data, city dashboards could show and track them and, in so doing, could support well-informed decisions by policy makers, urbanists, and architects.

\item \emph{Quantifying regulatory interventions.}  In his paper `A City is Not a Tree'~\cite{alexander1965}, Christopher Alexander argued that the variety and diversity essential for urban vitality was being destroyed by the implementation of zoning laws. Our work has focused on urban diversity features, the very  features that zoning laws impact. To track the impact of a regulation, one now knows \emph{which features} to contrast before and after the implementation of the regulation. 

\item \emph{Place recommendation.} How to best place retail stores? Researchers have recently tried to answer that question with semi-automatic mechanisms that identify the amenities that are missing from a neighborhood. Hidalgo \emph{et al.}~\cite{hidalgo2015we} analysed the economic diversity of neighborhoods  to recommend new places for them. Similar analyses could profit from the addition of  our work's structural features to the already present economic ones. 

\end{description}

\mbox{ } \\ 
\textbf{Theoretical Implications.} Some of Jane Jacobs's theories continue to be controversial, while others have been criticized for their non-verifiability. This study showed that those theories not only are verifiable but also are still important today and valid in the European context too. 








\mbox{ } \\ 
\textbf{Limitations.} Our measure of activity has been assumed to be  a good proxy for city's vitality. However, that measure has left out one important factor that might be extracted in the future: temporal dynamics of district activity  (e.g., differences  between day and night). Also, if one were to have individual calls records instead of (as in our case) aggregate ones, one could well extract additional factors: duration of individual stays, radius of gyration, and identification of particular segments of the population (e.g., pedestrians \emph{vs.} non-pedestrians, residents \emph{vs.} dwellers). Finally, it would be difficult to \emph{fully} replicate this study without call records. Those records were difficult to obtain in the past but have now been increasingly made publicly available by telecoms operators for research purposes~\cite{barlacchi2015}.

\section{Conclusion}
For the first time, we verified Jane Jacobs's four conditions necessary for the promotion of urban life in the Italian context. We have done so by  operationalizing her concepts in new ways: we used mobile phone records to extract a proxy for urban vitality, and web data to extract structural proxies for urban diversity. As Jacobs envisioned, vitality and diversity are intimately linked.

Until now, the research community has relied on limited data sets that reflect how a group of people experience bits of the city.  Here we have found that web data and mobile records offer  insights on how most urban dwellers experience entire cities. Web data (especially Open Street Map data) is available around the world, and mobile phone records have been increasingly made publicly available by telecoms operators in search of new business models. As such, our study can be replicated at scale without resorting to lengthy and costly survey-based data collection. 

By collecting bread crumbs from cell phones, social media, and participatory platforms, researchers will increasingly rely on data sets orders of magnitude richer than previous urban studies data sets~\cite{batty-big-data} and, consequently, they will be able to test traditional urban theories in fine-grained detail - a totally new way to look at cities.



\section{Acknowledgments}
We thank OpenStreetMap and all its contributors for making their data freely available.

\balance
%
\bibliographystyle{abbrv}
\bibliography{sigproc}  
%
%

\end{document}